\documentclass[sigconf, nonacm]{acmart}
\usepackage{booktabs}   
\usepackage{makecell}  
\usepackage{algorithm}
\usepackage{algpseudocode}
\usepackage{float}
\usepackage{tabularx}
\usepackage{placeins}
\usepackage{balance}   
\usepackage{flushend}  
\setcopyright{none}
\copyrightyear{2026}
\acmYear{2026}
\renewcommand\footnotetextcopyrightpermission[1]{}

\title{LLM-Derived Priors for Thompson Sampling in Cold-Start Comment Recommendation}

\author{Eugene Lee}
\authornote{Equal contribution.}
\email{eugenie@webtoonscorp.com}
\affiliation{%
  \institution{NAVER WEBTOON}
  \country{South Korea}
}

\author{Oseong Choi}
\authornotemark[1]
\email{oseong.choi@webtoonscorp.com}
\affiliation{%
  \institution{NAVER WEBTOON}
  \country{South Korea}
}

\author{Byungsoo Kang}
\email{bsoo414@webtoonscorp.com}
\affiliation{%
  \institution{NAVER WEBTOON}
  \country{South Korea}
}

\author{Taeyeong Jang}
\email{teo.jang@webtoonscorp.com}
\affiliation{%
  \institution{NAVER WEBTOON}
  \country{South Korea}
}

\begin{document}
\begin{abstract}

Multi-armed bandit algorithms, especially Thompson sampling, are widely used in online recommendation. Despite their ability to adapt from online feedback, these methods often suffer from cold-start limitations when newly introduced arms have little or no interaction history. In our setting, the candidate arms are user-generated textual comments, whose semantic content can reveal a title's appeal before sufficient interaction feedback is available. We therefore use large language models (LLMs) to extract semantic signals from comment text and convert them into informative Bayesian priors that warm-start Thompson sampling under sparse early-stage feedback. To account for aggregate segment-level differences in response patterns, we maintain and update posteriors separately for each gender--age segment. In a real-world online A/B/C test, we compare a uniform prior with two LLM-based designs: a Gender Prior for demographic-affinity cues and a Content Prior for title-specific identity cues. The results show that LLM-based priors are most beneficial in sparse-feedback regimes—with the largest gains emerging once a small amount of interaction evidence has accumulated—and that prior design leads to distinct funnel-level effects. We further analyze prior--reward alignment and demographic heterogeneity, finding that click-oriented alignment is strongest for the Gender Prior and that treatment effects vary substantially across demographic segments. These findings suggest that LLM-derived priors can serve as a practical warm-start mechanism for text-rich bandit recommendation, while also revealing deployment trade-offs.

\end{abstract}

\maketitle
\section{Introduction}

Personalized recommendation has become increasingly important in digital content platforms, where users must navigate large and diverse catalogs with highly individual tastes. As the number of available items grows, it becomes harder for users to identify content that matches their own preferences without algorithmic support. This challenge is particularly salient in the Webtoon domain. In the large-scale production platform considered in this paper, users face a catalog of roughly 4,000 active titles while differing widely in the kinds of works they prefer, including artistic style, narrative tone, character composition, pacing, and emotional themes. Identifying the right title therefore requires not only broad exposure to the catalog, but also sufficiently informative recommendation signals that can help users recognize works aligned with their interests.

This difficulty becomes especially pronounced at the discovery stage, where users make rapid decisions from thumbnails, title names, and limited metadata. These visual cues often provide only a partial explanation of what a work is actually like, so titles that are preference-matched by the upstream ranker may still be overlooked. To supplement this visual discovery surface, the platform introduced a comment recommendation component. User-generated comments can make a title's appeal more tangible by conveying tone, character dynamics, emotional tension, or distinctive thematic identity in the words of other readers.

When designing a recommendation policy for such a component, a key challenge is that the system is newly launched and therefore begins with sparse feedback. In this setting, multi-armed bandit methods are attractive because they support incremental online adaptation~\cite{li2010contextual,chapelle2011empirical,russo2018tutorial}. In particular, we maintain Thompson sampling parameters at the demographic-segment level rather than pooling all users together. This segment-level design reflects domain knowledge that user behavior and content preferences on the platform differ across gender and age groups, while avoiding the severe data fragmentation that would arise under fully user-specific posterior estimation. Section~4.5 later provides descriptive evidence of substantial heterogeneity in the effects of LLM-based priors across demographic segments. At the same time, sparse early-stage feedback means that even a segment-aware bandit can suffer from weak or largely unguided initialization, making prior construction a central problem in practical comment recommendation.

To address this issue, we propose an LLM-initialized, segment-aware Thompson sampling framework for comment recommendation. The key idea is to use the LLM's ability to extract semantic information from comment text in order to construct informative priors before sufficient behavioral feedback becomes available. We then evaluate two alternative prior designs---the Gender Prior and the Content Prior---under demographic-segment-level posterior updates in a real-world online deployment.

The main contributions of this work are as follows:
\begin{itemize}
    \item A segment-aware Thompson sampling framework that connects LLM-derived textual signals to Bayesian prior initialization with continual posterior updates from behavioral feedback.
    \item Two alternative LLM-based priors---the Gender Prior and the Content Prior---that capture distinct cue types and induce distinct effects across the recommendation funnel.
    \item A comprehensive evaluation of LLM-initialized bandit recommendation, including cold-start performance, prior--reward alignment, demographic heterogeneity, exposure concentration, and title bias in a real deployment setting.
\end{itemize}

\section{Related Work}
\subsection{Bandit-Based Recommendation}

Multi-armed bandit methods have been widely adopted in online recommendation because they address the exploration--exploitation trade-off under partial feedback and adapt sequentially to observed user responses~\cite{li2010contextual,chu2011contextual,lattimore2020bandit}. Among them, Thompson sampling is especially attractive for our setting because its Bayesian formulation provides a natural way to represent uncertainty, update beliefs incrementally, and incorporate prior knowledge about newly introduced arms~\cite{russo2018tutorial,chapelle2011empirical}.

Our work focuses on a practical issue in Bayesian bandit recommendation: how to initialize newly introduced arms when interaction history is scarce. Although prior specification is inherent to Thompson sampling, many practical systems still begin with weak or generic initialization when newly introduced items have little behavioral evidence. In our setting, this issue becomes a first-order design problem because many comments initially have little or no feedback, so the quality of the prior can strongly affect early-stage serving behavior.

Recent work has also begun to explore how LLM-derived knowledge can be used to warm-start bandit learning. Most closely related to our motivation, Alamdari et al.~\cite{alamdari2024jump} use LLM-generated preference data to jump-start contextual bandits and reduce early regret. Our work shares the goal of improving early-stage bandit learning with LLM-derived knowledge, but differs in both the bandit formulation and the deployment setting. Rather than pretraining a contextual bandit with synthetic preference data, we convert semantic signals extracted from user-generated comment text into Beta priors for Thompson sampling and evaluate the resulting policy in a real-world online comment recommendation system.

\subsection{Textual Signals for Cold-Start Recommendation}

Previous studies have addressed cold-start and sparsity problems by exploiting textual signals associated with users or items, such as item descriptions and user-generated reviews. These signals are commonly used to enrich item or user representations, or to improve prediction quality when behavioral feedback is limited. Representative examples include Collaborative Deep Learning, which jointly learns from item content and user--item feedback~\cite{wang2015cdl}, and DeepCoNN, which derives user and item representations from reviews~\cite{zheng2017deepconn}. Review text can contain preference-relevant signals beyond ratings~\cite{mcauley2013hidden}, although such signals are not uniformly useful and must be incorporated carefully~\cite{sachdeva2020useful}.

Our setting is related in that we also use text under sparse feedback, but the role of text is different: the recommended arms themselves are user-generated comments. We use each comment's text directly to construct Bayesian priors for Thompson sampling, so that text influences early-stage online serving rather than only an offline representation or prediction model.

\subsection{LLMs for Recommendation and Ranking}

Recent work has explored the use of large language models in recommendation and ranking, motivated by their ability to infer semantic and preference-related signals from natural language. Representative approaches include reformulating recommendation as a language-processing problem~\cite{geng2022recommendation}, using LLMs for zero-shot or prompt-based recommendation~\cite{hou2023zeroshot}, and applying LLMs as rankers or rerankers over candidate items~\cite{qin2023pairwise}. Recent surveys provide broader taxonomies of LLM-based recommendation methods~\cite{wu2024survey}.

Our work is related to this line of research, but differs in how the LLM output is incorporated into the recommendation policy. Rather than using the LLM as a standalone recommender, generator, or reranker at serving time, we use it as an offline semantic prior estimator. This integration is natural because Thompson sampling already requires prior specification, while LLMs can infer informative semantic signals directly from text. Using the LLM to construct priors therefore complements online behavioral learning, rather than replacing the bandit policy with a standalone language-model ranker. The resulting prior signals are converted into Beta priors and updated online through Thompson sampling with impression and click feedback.

\section{Method}

The proposed framework operates in two stages. When a comment first becomes eligible for recommendation, the system constructs an informative prior from its text using an LLM. This prior is built in an additive form: a shared base score captures broadly appealing hook strength, and two alternative adjustment modules---the Gender Prior and the Content Prior---modify this base according to demographic-affinity cues or title-specific identity cues, respectively. The resulting prior is then converted into Beta pseudo-counts for Thompson sampling.

Once the comment begins receiving impressions and clicks, these interaction signals are accumulated at the demographic-segment level and used to update posterior parameters on an hourly basis. In this way, the model combines an LLM-derived cold-start prior with online behavioral feedback, allowing newly introduced comments to be served before sufficient interaction history has accumulated while still adapting quickly as evidence grows. The remainder of this section formalizes the recommendation problem, the prior construction procedure, the posterior update mechanism, and the overall offline--online pipeline.

\subsection{Problem Setup}
For each request, let $\mathcal{A}$ denote the candidate arm set. Candidate comments are not drawn from the full global comment inventory, but from a user-specific title pool preselected by an upstream title ranker. Each arm $i \in \mathcal{A}$ corresponds to a candidate comment.

Each request is associated with a segment
\begin{equation}
s = (g, a),
\end{equation}
where $g \in \{\texttt{F}, \texttt{M}\}$ denotes gender, and age is grouped into four operational segments: \(\leq 17\), 18--27, 28--37, and 38+. These bins follow the platform's standard demographic segmentation.

For a displayed comment $i$ and segment $s$, the reward is defined as a Bernoulli click event observed after the comment preview is shown:
\begin{equation}
r_{i,s} = \mathbb{I}\{\text{click on comment } i \text{ by a user in segment } s\}.
\end{equation}
Downstream reading behavior, which we report as CVR in Section~4, is treated as an auxiliary outcome and is not used as the bandit reward.

We model the corresponding click probability as
\begin{equation}
\theta_{i,s} \sim \mathrm{Beta}(\alpha_{i,s}, \beta_{i,s}).
\end{equation}
At serving time, one posterior sample is drawn for each candidate comment,
\begin{equation}
\tilde{\theta}_{i,s} \sim \mathrm{Beta}(\alpha_{i,s}, \beta_{i,s}),
\end{equation}
and the \(K\) comments with the largest sampled values are selected for display:
\begin{equation}
\mathcal{R} = \operatorname{Top-K}_{i \in \mathcal{A}} \tilde{\theta}_{i,s},
\end{equation}
where $K=10$ in our deployment.

\subsection{LLM-Based Prior Construction}

The key role of the LLM in our framework is to provide informative priors before sufficient behavioral feedback becomes available. Early prompt development suggested that comment appeal in this setting comes from multiple sources. Some comments are broadly attractive to first-time readers, while others are useful because they express title-specific identity or align with segment-level response patterns. This observation motivated a decomposed prior structure: the model first estimates a shared base score for broadly appealing comments, and then applies either a gender-specific or a content-specific adjustment.

This additive design has two practical advantages. First, it separates general hook strength from more specific adjustment signals, making the resulting priors easier to inspect and debug. Second, it avoids estimating fully separate priors from scratch for every prior type, which made the outputs more stable and operationally reliable in our prompt development. The resulting adjusted priors are evaluated separately in the main online experiment.

\subsubsection{Prompt Design for LLM Scoring}

The prompt design follows this decomposed structure. The base score prompt estimates a click-through-rate-like score for each comment, while the gender and content prompts estimate task-specific adjustment scores. We ask the model to produce CTR-like quantities rather than abstract relevance scores so that the outputs can be used directly as prior means or prior adjustments without an additional score-to-probability calibration step.

LLM scoring is implemented as a title-level batch-scoring task. In each call, the model receives the candidate comments for a title together with unique comment identifiers and returns structured JSON outputs keyed by those identifiers. The identifiers are used only to merge the model outputs back to the corresponding comments. This title-level format allows the model to use surrounding comment context, which is particularly important for content analysis, while keeping the outputs easy to validate and merge into downstream prior artifacts.

\paragraph{LLM scoring configuration}
All scoring modules use GPT-4.1. We set temperature to 0.0 for base score, 0.1 for title-level content analysis, and 0.3 for content- and gender-delta scoring. All input comments are written in Korean, the native language of the service. Each title-level call takes a list of candidate comments with unique identifiers and returns JSON records keyed by those identifiers. A simplified prompt template and module-level output schema are provided in Appendix~\ref{app:prompt_template}.

\paragraph{Base score}
The base score prompt estimates a universal hook-strength score, capturing comments that are generally attractive to first-time readers independent of title-specific identity or demographic preference. The prompt favors comments that are concise, immediately understandable, and affectively engaging, while downweighting comments that are generic, overly descriptive, diary-like, or dependent on insider context.

The score is calibrated as a CTR-like value in $[0,1]$, with most comments expected to fall below $0.50$. Very weak comments fall in $[0.00,0.10]$, typical comments in $[0.10,0.30]$, and strongly hooking comments in $[0.30,0.50]$, while scores above $0.50$ are reserved for rare exceptional cases.

\paragraph{Gender-delta prompt}
The gender adjustment is designed to capture aggregate segment-level differences in response to comment phrasing, tone, and expressed appeal. We use demographic information only as a weak signal for cold-start prior calibration, not as a fixed assumption about individual users.

The prompt is calibrated to keep gender effects bounded and interpretable. Comments with no clear segment-specific signal are assigned values close to zero, typically within $[-0.02, 0.02]$. Positive affinity signals receive bonuses in $[0.03, 0.30]$, with rare strong bonuses up to $0.40$, while aversion signals receive penalties in $[-0.20, -0.05]$, with rare strong penalties down to $-0.35$. This gives an intended adjustment range of approximately $\Delta_i^F, \Delta_i^M \in [-0.35, 0.40]$ before the weighted adjustment is applied.

\paragraph{Content-delta prompt}
The content adjustment gives more exposure to comments that make a title's distinctive appeal legible to prospective readers. We use a two-stage design. First, a title-level prompt reads the full eligible comment set of a title and extracts five representative themes or strengths. These themes are inferred from reader responses rather than drawn from a fixed taxonomy, allowing them to reflect the aspects of the work most salient in the comment set. We use five themes to preserve a compact representation of title identity.

A second prompt then scores which comments best express one or more of these inferred themes. The content adjustment is calibrated as a non-negative additive score, $\Delta_i^T \in [0,1]$, so it can increase the base prior for comments that clearly express title-specific appeal without penalizing comments that lack such signals.

\subsubsection{Prior Estimation from Comment Text}

Let $b_i \in [0,1]$ denote the base score for comment $i$, and let
$\Delta_i^F,\Delta_i^M \in [-0.35,0.40]$ and
$\Delta_i^T\in[0,1]$ denote the gender and content (title-identity) adjustment terms produced by the LLM prompts, respectively. We combine these outputs into two alternative prior means. For the Gender Prior, the prior mean for comment $i$ and segment $s$ is
\begin{equation}
\mu_{i,s}^{G} =
\operatorname{clip}\!\left(
b_i + \lambda_G\Delta_i^{g(s)},\,0,\,1
\right),
\end{equation}
where $\lambda_G, \lambda_T \geq 0$ are scaling coefficients for the gender 
and content adjustments, respectively, and $g(s)\in\{\texttt{F},\texttt{M}\}$ 
denotes the gender associated with segment $s$. For the Content Prior, the prior mean is
\begin{equation}
\mu_i^{T} =
\operatorname{clip}\!\left(
b_i + \lambda_T\Delta_i^T,\,0,\,1
\right).
\end{equation}
In our implementation, we set $\lambda_T=1.0$ and $\lambda_G=2.0$ through qualitative calibration rather than exhaustive optimization.

\subsubsection{Conversion to Beta Priors}

Each prior mean is converted into Beta pseudo-counts. Let $\kappa$ denote the prior strength, interpreted as the number of pseudo-impressions. For a prior mean $\mu$, we define
\begin{align}
\alpha_0 &= 1 + \operatorname{round}(\kappa\mu),\\
\beta_0  &= 1 + \operatorname{round}(\kappa(1-\mu)).
\end{align}
The additive constant $1$ prevents degenerate zero-count cases. We use a fixed prior strength $\kappa=40$ for all comments. This value was chosen as an experimental setting that makes the effect of LLM-derived priors observable under sparse traffic, rather than as a production-optimized hyperparameter. In our traffic logs, the median number of 7-day impressions per comment was 12 and the 75th percentile was approximately 40. Since the relative weight of the prior after $n$ impressions is roughly $\kappa/(\kappa+n)$, setting $\kappa=40$ keeps the prior influential for most comments while allowing behavioral evidence to match the prior weight once a comment reaches the upper quartile of exposure. 

The three experimental variants instantiate this conversion differently. Variant~A uses a uniform prior, $\alpha_{0,i,s}^{U}=\beta_{0,i,s}^{U}=1$. For Variants~B and~C, the conversion is applied with $\mu=\mu_{i,s}^{G}$ and $\mu=\mu_i^{T}$, respectively. Since the Content Prior is not demographic-specific, the same prior mean is used across segments, while posterior updates remain segment-specific.

\subsection{Segment-Aware Thompson Sampling}

Posterior updates are performed hourly using recent interaction logs. For each comment $i$ and segment $s$, we aggregate the number of impressions and clicks observed over the most recent 7-day window:
\begin{equation}
n_{i,s} = \sum_{\tau \in \mathcal{W}} \mathrm{imp}_{i,s,\tau},
\qquad
k_{i,s} = \sum_{\tau \in \mathcal{W}} \mathrm{click}_{i,s,\tau},
\end{equation}
where \(\mathcal{W}\) denotes the 7-day update window, \(n_{i,s}\) is the total number of impressions, and \(k_{i,s}\) is the total number of clicks for comment \(i\) in segment \(s\).

These counts are combined with the prior parameters to form the posterior:
\begin{align}
\alpha_{i,s} &= \alpha_{0,i,s} + k_{i,s},
\label{eq:posterior_alpha}\\
\beta_{i,s}  &= \beta_{0,i,s} + n_{i,s} - k_{i,s}.
\label{eq:posterior_beta}
\end{align}
Intuitively, clicks increase the success count, while impressions without clicks increase the failure count.

At serving time, the system looks up the posterior corresponding to the user's segment, draws one posterior sample for each candidate comment, and displays the \(K\) comments with the largest sampled values.

\subsection{Offline--Online Update Pipeline}

The overall system follows an offline--online pipeline operating at two timescales. In the offline stage, the LLM-based prior construction pipeline is executed before deployment for all eligible titles and comments. After deployment, the pipeline is rerun daily only for titles or comments that are newly added or changed relative to the previous run. The resulting scores are merged with previously saved prior artifacts, allowing the system to amortize the cost of semantic inference while keeping the prior table up to date. This cadence is sufficient in our setting because the eligible comment pool changes incrementally rather than in real time.

To ensure that incomplete LLM outputs do not silently affect serving, the offline pipeline validates the generated prior table against the latest metadata at both the title and comment levels. Validation flags titles missing from the prior output or with $>1\%$ comment-level coverage loss. If such cases are detected, the pipeline logs sample title identifiers and fails before producing the final serving artifact; the previously deployed prior artifact remains in use until the issue is resolved. Comments that are no longer valid according to the latest metadata are removed before posterior construction.

In the online stage, interaction logs are aggregated hourly and combined with the daily prior artifacts to refresh posterior parameters. We use a 7-day sliding window for posterior construction. The two stages interact through the posterior updates in Equations~\eqref{eq:posterior_alpha}--\eqref{eq:posterior_beta}, where the daily prior parameters are combined with recent impression and click counts to form the final segment-specific posterior. This separation allows the system to perform expensive LLM-based semantic scoring offline while still adapting quickly to recent user feedback. A compact algorithmic summary of the offline prior construction, posterior update, and serving procedure is provided in Appendix~\ref{app:algorithm}.

\section{Evaluation}
The proposed method is evaluated from four complementary perspectives. The analysis begins with overall online performance, using CTR and CVR as the primary metrics to assess how the prior variants affect different stages of the interaction funnel. Cold-start performance is then examined to test whether LLM-based priors provide value when interaction history is sparse. Next, prior--reward alignment is studied to assess whether the prior scores themselves are directionally aligned with observed reward. Finally, demographic heterogeneity is analyzed to examine whether the effects of the priors differ systematically across gender--age segments.

\subsection{Experimental Setup}

An online A/B/C test was conducted on the comment recommendation component. Users were assigned to one of three treatment variants that differ only in the prior initialization strategy: Variant~A uses a uniform prior and serves as the control benchmark, Variant B uses the Gender Prior, and Variant C uses the Content Prior. Traffic was approximately balanced across the three variants, with about 595K users assigned to each condition. The main analysis uses a fixed four-week window from March 7 to April 3, 2026, rather than the full deployment period, in order to preserve a consistent early-stage comparison across variants.

Comment ranking is not performed over the full global comment inventory. Instead, for each user, candidate comments are drawn only from a constrained title pool defined by the upstream new-title recommendation model. In the deployed system, this pool consists of the 50 highest-ranked titles for that user, from which the comment recommendation model selects 10 comments for display. Because the upstream title pool can change over time, the comment bandit operates on a dynamic candidate set rather than a fixed global inventory.

As this was a newly launched component, the candidate comment pool was manually curated before recommendation to mitigate operational risk. Only comments that passed this manual screening process were eligible for serving. This makes the evaluation conservative, because even the control condition starts from a relatively strong candidate pool rather than from the full set of noisy comments.

During the analysis period, the component covered approximately 1.0K--1.3K titles out of roughly 4K active titles. The eligible comment pool expanded from 18.8K to 22.0K comments, with a median of 16 and a mean of 17 candidate comments per title, as manual comment screening and selection continued alongside the online experiment.

We report two primary online metrics. The first is click-through rate (CTR), which measures immediate response at exposure time. For a set of impressions $\mathcal{I}$, we define
\begin{equation}
\mathrm{CTR}(\mathcal{I}) = \frac{\sum_{e \in \mathcal{I}} \mathrm{click}_e}{|\mathcal{I}|},
\end{equation}
where $\mathrm{click}_e \in \{0,1\}$ indicates whether the user clicked the comment preview for impression $e$.

The second metric is conversion rate (CVR), which measures downstream reading conditional on click. Let $\mathcal{J}$ denote the set of clicked impressions. We define
\begin{equation}
\mathrm{CVR}(\mathcal{J}) = \frac{\sum_{e \in \mathcal{J}} \mathrm{read}_e}{|\mathcal{J}|},
\end{equation}
where $\mathrm{read}_e \in \{0,1\}$ indicates whether the click on impression $e$ was followed by downstream reading behavior within the attribution window.

For each treatment variant $v \in \{A, B, C\}$, we report the relative lift of a metric $M \in \{\mathrm{CTR}, \mathrm{CVR}\}$ against the control variant $A$:
\begin{equation}
\mathrm{Lift}_M(v \mid A) =
\frac{M(v) - M(A)}{M(A)} \times 100.
\end{equation}

CTR and CVR capture different stages of the interaction funnel and are therefore both reported in the overall online comparison. However, the subsequent analyses emphasize CTR, since the proposed priors are explicitly designed to initialize click-oriented Thompson sampling parameters, whereas CVR is treated as a complementary downstream metric.

Throughout the evaluation section, confidence intervals are 95\% percentile intervals from a cluster bootstrap ($B = 1000$ replications). For overall and segment-level analyses we cluster at the user level (the unit of A/B/C randomization); the cold-start analysis (Section~4.3) clusters at the comment level since it conditions on comment-level exposure history. Significance is assessed via a two-sided bootstrap sign test against Variant~A.

\subsection{Overall Performance}

\begin{table}[t]
\centering
\caption{Overall online CTR and CVR across the three prior configurations.
Point estimates and 95\% confidence intervals (brackets) are from a
user-level cluster bootstrap ($B = 1000$ replications). Lift is reported
relative to Variant A; significance is from a two-sided bootstrap
sign test against A.}
\label{tab:metrics_summary}
\centering
\setlength{\tabcolsep}{4pt}
\small
\begin{tabular}{lcccc}
\toprule
Variant & CTR (\%) & CTR Lift (\%) & CVR (\%) & CVR Lift (\%) \\
\midrule
A (Uniform)
  & \makecell{1.819 \\ {\scriptsize [1.795,\,1.845]}}
  & \textemdash
  & \makecell{41.095 \\ {\scriptsize [40.721,\,41.484]}}
  & \textemdash \\
\addlinespace
B (Gender)
  & \makecell{1.846 \\ {\scriptsize [1.822,\,1.872]}}
  & \makecell{$+$1.48 \\ {\scriptsize [$-$0.48,\,$+$3.55]}}
  & \makecell{41.599 \\ {\scriptsize [41.231,\,41.995]}}
  & \makecell{$+$1.23 \\ {\scriptsize [$-$0.15,\,$+$2.54]}} \\
\addlinespace
C (Content)
  & \makecell{1.716 \\ {\scriptsize [1.691,\,1.739]}}
  & \makecell{$-$5.68$^{***}$ \\ {\scriptsize [$-$7.71,\,$-$3.80]}}
  & \makecell{41.656 \\ {\scriptsize [41.277,\,42.098]}}
  & \makecell{$+$1.37 \\ {\scriptsize [$-$0.04,\,$+$2.77]}} \\
\bottomrule
\end{tabular}
\\[3pt]
{\footnotesize $^{*}\,p<0.05$, $^{**}\,p<0.01$, $^{***}\,p<0.001$. The Gender Prior's overall CTR/CVR gains and the Content Prior's CVR gain are directionally positive but not significant at $p<0.05$ at the aggregate level; subgroup analyses in Sections~4.3 and~4.5 show statistically significant effects in sparse-feedback buckets and selected demographic segments.}
\end{table}

Table~\ref{tab:metrics_summary} summarizes the aggregate performance over the analysis period. Variant~B (Gender Prior) shows a directional CTR gain of $+1.48\%$ over the control and a CVR gain of $+1.23\%$, although neither is statistically significant at the aggregate level under user-level cluster bootstrap ($p = 0.144$ and $p = 0.082$, respectively). Variant~C (Content Prior) significantly lowers CTR by $-5.68\%$ ($p < 0.001$) but yields a directionally positive CVR gain of $+1.37\%$ ($p = 0.054$).

Despite the muted aggregate picture, the two priors appear to affect different stages of the funnel: the Gender Prior is directionally aligned with immediate click behavior, whereas the Content Prior is less click-oriented and more aligned with downstream conversion. Subsequent analyses in Section~4.3 and Section~4.5 show that both effects become substantially stronger and statistically significant once the population is sliced by feedback sparsity or demographic segment. This funnel-stage difference is also consistent with the intended design of the two priors: demographic-affinity cues are expected to be more immediately legible, while title-specific identity cues act as a self-selection filter---clicking a content-rich comment implicitly signals receptivity
to a specific appeal of the work, so the Content Prior's lower CTR and
higher CVR reflect a single mechanism rather than two unrelated effects.

\subsection{Cold-Start Performance}
\label{sec:coldstart}
Cold-start is a primary motivation for LLM-based priors: when a comment has little interaction history, the Thompson-sampling posterior is weakly informed by behavioral feedback and the prior dominates serving decisions. We therefore evaluate whether the two LLM-based priors provide larger benefits than the uniform-prior baseline in sparse-feedback regimes.

\begin{figure}[t]
    \centering
    \includegraphics[width=\linewidth]{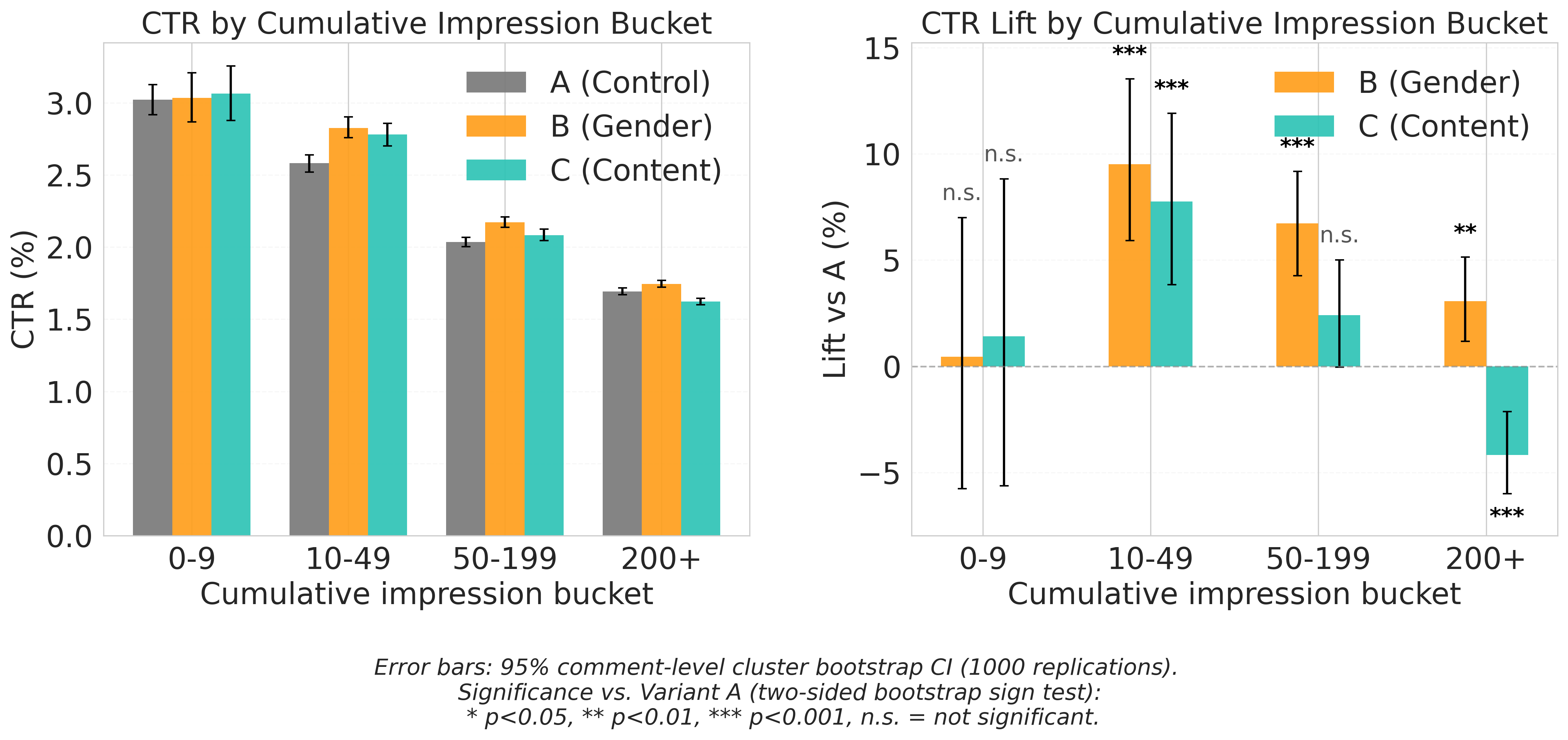}
    \caption{CTR and CTR lift by cumulative impression bucket. For each impression, comments are grouped according to the lagged cumulative number of prior impressions observed before that exposure. The left panel reports observed CTR for each variant, and the right panel reports CTR lift of Variant~B (Gender Prior) and Variant~C (Content Prior) relative to Variant~A (Uniform Prior). Both LLM-based priors show their largest lifts in the 10--49 bucket (both $p < 0.001$), with the advantage attenuating as cumulative exposure grows. Error bars denote 95\% comment-level cluster bootstrap CIs ($B = 1000$); significance markers in the right panel correspond to a two-sided sign test against Variant~A ($^{**}\,p<0.01$, $^{***}\,p<0.001$; n.s.\ = not significant).}
    \label{fig:ctr_cumulative_impression_bucket}
\end{figure}

To measure comment-level feedback sparsity more directly, we group impression events by the cumulative number of prior impressions already accumulated by the same comment. In other words, each exposure is assigned to a bucket according to how many times that comment had been shown before the current exposure occurred. We use four cumulative impression buckets: \texttt{0--9}, \texttt{10--49}, \texttt{50--199}, and \texttt{200+}. This analysis is therefore impression-level rather than comment-level: the same comment may contribute to different buckets over time as its cumulative exposure grows. Figure~\ref{fig:ctr_cumulative_impression_bucket} reports the resulting CTR levels and the corresponding CTR lift relative to Variant~A.

In the coldest 0–9 bucket, lifts are directionally positive but not statistically significant, partly reflecting the wide confidence intervals induced by limited per-comment impressions in this regime; the pattern across buckets suggests that the priors act primarily as accelerators of early-stage learning once minimal feedback arrives, rather than as a substitute for initial exploration. In the 10--49 bucket, the Gender Prior improves CTR by +9.51\% (95\% CI [+5.92,+13.53], $p<0.001$) and the Content Prior by +7.76\% ([+3.84,+11.91], $p<0.001$). In the 50--199 bucket, the Gender Prior maintains a +6.72\% ([+4.26,+9.18], $p<0.001$) improvement. The advantage persists into the 200+ bucket for the Gender Prior ($+3.07\%$, $p < 0.01$) but reverses for the Content Prior ($-4.16\%$, $p < 0.001$), suggesting that the Content Prior is more useful as an early title-identity signal than as a consistently click-oriented ranking signal.

\subsection{Prior--Reward Alignment}

While the overall online results indicate whether a prior improves CTR or CVR at the policy level, they do not directly show whether higher prior scores correspond to better-performing comments. We therefore analyze whether each prior induces a meaningful ranking over comments under a less policy-contaminated benchmark.

To do so, we group comments into deciles according to their prior score and compare them against observed CTR measured under Variant~A (Uniform Prior). We use Variant~A as the benchmark because the two LLM-based treatment variants can alter exposure allocation in a self-reinforcing way, making direct post-treatment calibration against their own observed CTR less informative. Since the prior scores are not necessarily calibrated probabilities, we focus on rank alignment and monotonicity rather than absolute probability calibration. In addition to the universal base score and the Content Prior, we evaluate the Gender Prior separately for female- and male-specific scores.

\begin{figure}
    \centering
    \includegraphics[width=0.9\linewidth]{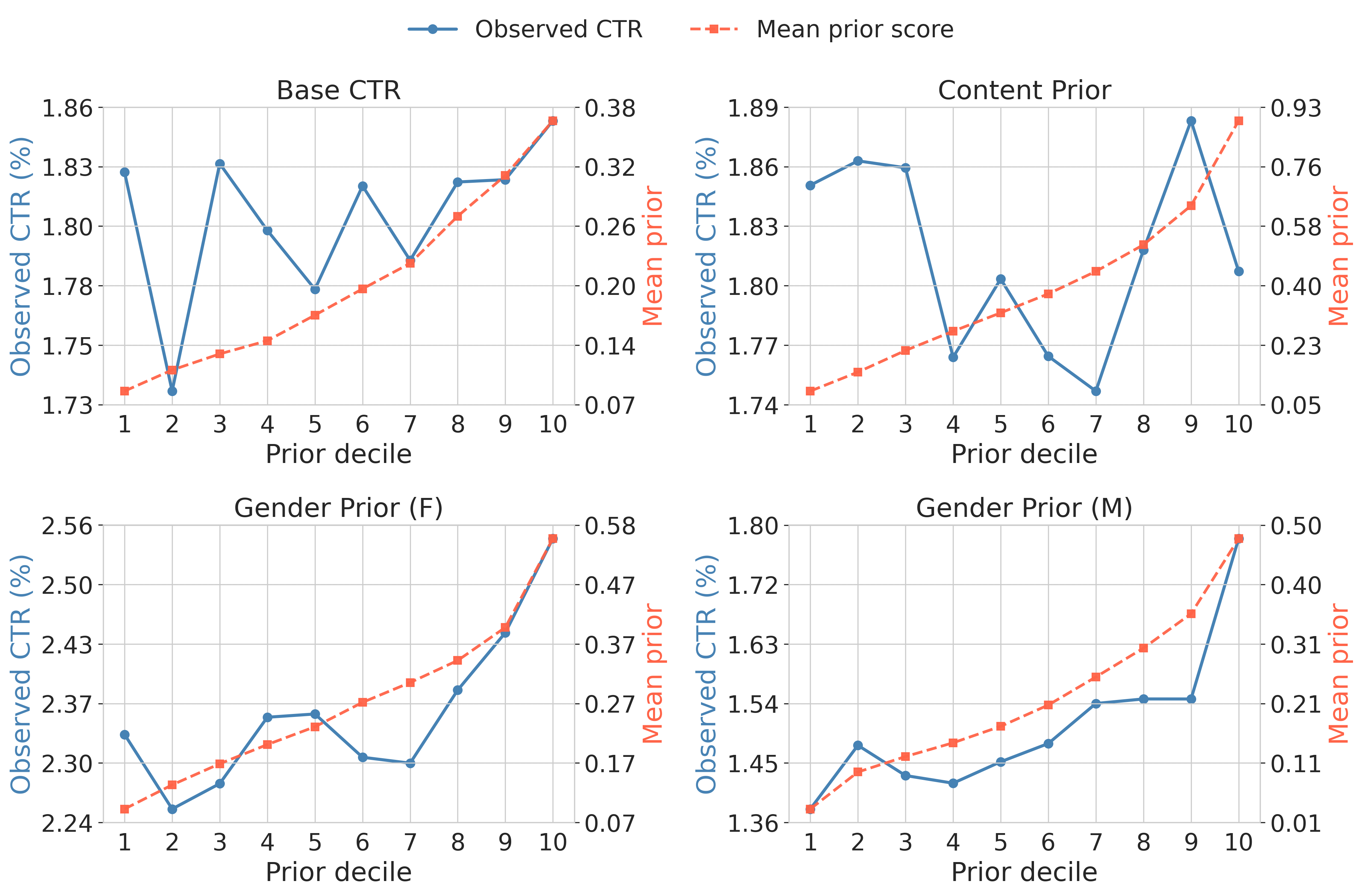}
    \caption{CTR prior quality under the Variant~A benchmark. Comments are grouped into deciles according to each prior score, and the corresponding observed CTR under Variant~A is plotted against the mean prior score in each decile. The two gender-specific priors exhibit the clearest positive alignment with observed CTR, whereas the base score shows only weak alignment and the Content Prior does not exhibit a clear monotonic relationship with CTR.}
    \label{fig:prior_quality_ctr}
\end{figure}

Figure~\ref{fig:prior_quality_ctr} shows that the gender-specific priors
have the clearest positive alignment with observed CTR, while the base
score alone shows only weak alignment. The contrast supports the
additive design: the gender adjustment $\Delta_i^{F/M}$, layered on the
demographic-agnostic base score, surfaces more clickable comments. The
Content Prior does not exhibit a clear monotonic relationship with CTR,
suggesting that title-identity cues capture a different mechanism not
reflected in comment-level CTR monotonicity.

\subsection{Demographic Heterogeneity}
\label{sec:segment}
We next examine whether treatment effects vary across gender--age segments. This analysis is descriptive: it does not isolate the causal benefit of segment-specific posteriors over pooled posteriors, but it shows whether pooling would obscure meaningful response differences.

\begin{figure}[t]
    \centering
    \includegraphics[width=0.9\linewidth]{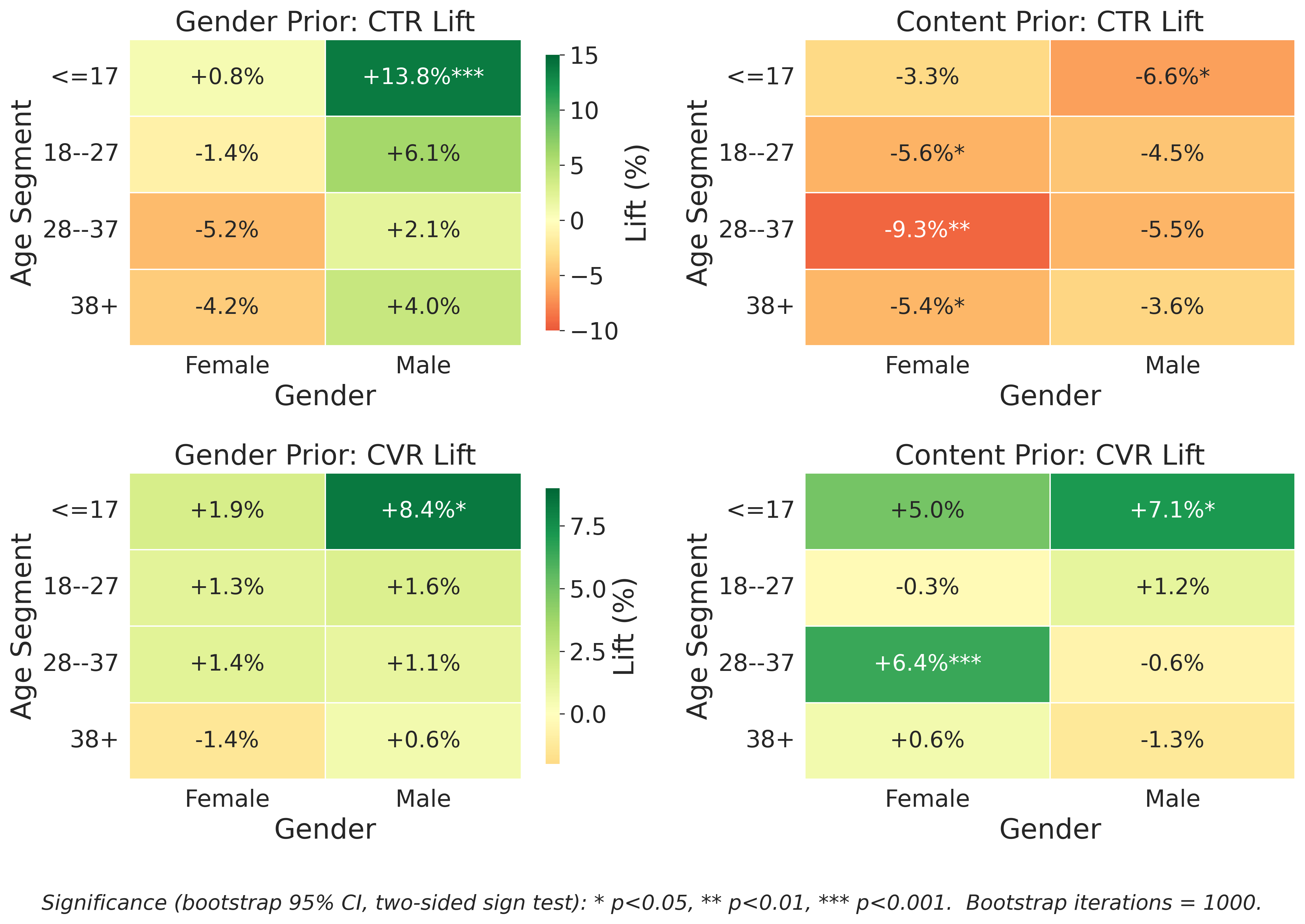}
    \caption{Segment-level CTR and CVR lift relative to Variant A (Uniform Prior). The heatmaps report treatment lift across the gender--age grid for Variant B (Gender Prior) and Variant C (Content Prior). Cell annotations report point estimates and significance markers from a user-level cluster bootstrap
($B=1000$): $^{*}\,p<0.05$, $^{**}\,p<0.01$, $^{***}\,p<0.001$ from a
two-sided sign test against Variant A. The results show substantial
heterogeneity across demographic segments, with segment-specific differences in both the magnitude and direction of treatment effects.}
    \label{fig:segment_heatmap}
\end{figure}

Figure~\ref{fig:segment_heatmap} shows that user response varies not only in magnitude but also in direction across the gender--age grid. For the Gender Prior, the CTR gains are concentrated in male segments, with the strongest improvement in males aged $\leq 17$ ($+13.76\%$ [$+7.16$, $+20.60$], $p < 0.001$); female segments show no significant CTR effect, with point estimates ranging from $-5.2\%$ to $+0.8\%$. The Content Prior produces statistically significant CTR declines for female segments aged 18--37 ($-5.6\%$ to $-9.3\%$, all $p < 0.05$), but at the same time a significant CVR gain for females aged 28--37 ($+6.42\%$ [$+2.66$, $+10.47$], $p < 0.001$) and  a significant CVR gain for males aged $\leq 17$ as well ($+7.13\%$, $p < 0.05$). These cell-level patterns confirm that the same prior can induce heterogeneous and even opposite-signed effects across demographic groups, and they refine the aggregate picture in two specific ways: the Gender Prior's CTR gain is driven primarily by male segments, while the Content Prior operates as a selective filter that reduces broad click propensity but increases conditional conversion in a smaller set of demographic groups.

These results confirm that pooling across demographic segments would obscure substantial response heterogeneity, supporting the use of segment-aware posterior updates to preserve such differences.

\section{Discussion and Limitations}
\label{sec:discussion}

The aggregate CTR and CVR effects under the LLM-based priors are modest, but the bucket- and segment-level analyses reveal that the gains are concentrated where each prior is designed to help---in sparse-feedback regimes (Section~4.3) and in segments aligned with the prior's signal (Section~4.5). The fact that the Content Prior's CTR decline is statistically robust while its CVR gain is directionally positive (though marginal at the aggregate level) is consistent with the interpretation that the Content Prior trades immediate click propensity for downstream conversion, rather than uniformly improving the funnel. To better understand why the LLM-based gains do not translate into uniformly stronger aggregate metrics, we next analyze two deployment factors that shape the observed effects of comment-level priors: exposure concentration and title-level confounding.

First, stronger priors can induce more exploitative serving. Both LLM-prior variants exposed fewer unique comments and concentrated impressions on a smaller set of comments than the uniform-prior baseline. This suggests that prior-based exploitation may trade off against exposure diversity and rotation, which can matter in lightweight browsing surfaces where freshness is important~\cite{kaminskas2016diversity,lathia2010temporal}. Detailed concentration metrics are reported in Appendix~\ref{app:exposure_concentration}.

Second, title-level attractiveness can attenuate the observable contribution of comment-level priors. In a diagnostic analysis, title identity alone explained a substantial share of comment CTR variance, ranging from 22\% to 51\% depending on the minimum impression threshold. Similar patterns were already present under the uniform-prior baseline, suggesting that this clustering is largely behavioral rather than induced by the LLM priors themselves. This interpretation aligns with prior work showing that click-based feedback can be shaped by popularity-related exposure effects, making it difficult to isolate the contribution of comment-level priors from title-level attractiveness~\cite{abdollahpouri2019unfairness,klimashevskaia2024popbias}. A supplementary popularity split, reported in Appendix~\ref{app:popularity_heterogeneity}, further suggests that the Gender Prior has larger relative gains for unpopular titles, where title-level attraction is weaker.

Several limitations should be noted. First, the experiment was conducted on a manually screened candidate pool rather than the full raw comment inventory, so the observed treatment effects should be interpreted as conservative. Second, the evaluation is tied to a specific discovery surface in which comments are displayed alongside salient title-level visual cues, and the results may differ in interfaces where title-level presentation signals are weaker~\cite{joachims2005accurately}. Third, we do not include an explicit ablation comparing segment-specific posterior updates against pooled posterior updates under the same LLM prior. Finally, prior--reward alignment is clearer for CTR than for CVR, indicating that comment-level monotonicity is a stronger diagnostic for click-oriented priors than for downstream conditional outcomes. Together, these limitations bound rather than overturn the present findings: the convergent evidence across cold-start, segment-level, and popularity analyses supports the framework's practical value, and the listed gaps mark concrete targets for follow-up deployment studies.
\FloatBarrier

\section{Conclusion and Future Work}
Our LLM-initialized, segment-aware Thompson sampling framework delivered its strongest gains under sparse feedback, with different prior designs producing distinct funnel-level effects and heterogeneous responses across demographic segments. More broadly, our results suggest that LLM-derived signals need not act as standalone rankers or generators: they can also serve as offline priors that warm-start an online bandit while leaving behavioral adaptation to the policy itself. We see this separation of expensive semantic inference from low-latency serving as a practical pattern for integrating LLMs into production recommendation stacks.

Promising next steps include disentangling title-level attractiveness from comment-level effects, scaling beyond a manually screened candidate pool, and adding diversity- or rotation-aware serving constraints to counter the exposure concentration observed under stronger priors.
\bibliographystyle{ACM-Reference-Format}
\bibliography{references}

\appendix

\section{Prompt Template}
\label{app:prompt_template}

The production prompts contain service-specific calibration details and are not reproduced verbatim. Instead, we provide a simplified template that summarizes the structured input--output format shared across LLM scoring modules. All modules are implemented as title-level batch-scoring tasks: the input contains task instructions, calibration guidance, and a list of comments represented by a unique key and raw Korean text; the output is a JSON object keyed by comment identifier, with one
record per input comment.

\begin{table}[H]
\centering
\caption{Summary of LLM scoring modules.}
\label{tab:prompt_modules}
\small
\setlength{\tabcolsep}{3pt}
\begin{tabularx}{\columnwidth}{l X l}
\toprule
\textbf{Module} & \textbf{Output field} & \textbf{Score range} \\
\midrule
Base score & \texttt{base\_score} & $[0,1]$ \\
Content analysis & \texttt{keywords} & Top-5 themes \\
Content delta & \texttt{content\_delta} & $[0,1]$ \\
Gender delta & \texttt{male\_delta}, \texttt{female\_delta} & $[-0.35,0.40]$ \\
\bottomrule
\end{tabularx}
\end{table}

\begin{table}[H]
\centering
\caption{Simplified structured prompt template.}
\label{tab:prompt_template}
\begin{tabularx}{\columnwidth}{p{0.34\columnwidth}X}
\toprule
\textbf{Component} & \textbf{Simplified template} \\
\midrule
System instruction & Return JSON only, matching the specified schema. \\
Task instruction & Score each candidate comment according to the target module. \\
Input unit & Title-level batch of eligible comments. \\
Comment format & Each comment is represented by a unique key and raw comment text. \\
Display constraint & Evaluate the comment preview using only the visible prefix of the comment. \\
Calibration & Use module-specific score anchors or adjustment ranges. \\
Output format & JSON only, with one record per input comment and task-specific score fields. \\
Downstream use & Parsed scores are merged into prior-construction artifacts. \\
\bottomrule
\end{tabularx}
\end{table}

\section{Algorithmic Summary}
\label{app:algorithm}

\begin{algorithm}[H]
\caption{Offline--online LLM-initialized Thompson sampling}
\label{alg:offline_online_ts}
\begin{algorithmic}[1]
\Require Candidate comments, segment set $\mathcal{S}$, prior strength $\kappa$
\Statex \textbf{Offline prior construction}
\State Score each title-level comment batch with the LLM
\State Obtain $b_i$, $\Delta_i^F$, $\Delta_i^M$, and $\Delta_i^T$
\State Construct prior means $\mu_{i,s}^{G}$ and $\mu_i^{T}$
\State Convert each prior mean $\mu$ into Beta pseudo-counts:
\[
\alpha_0=1+\operatorname{round}(\kappa\mu), \quad
\beta_0=1+\operatorname{round}(\kappa(1-\mu))
\]
\State Validate and save prior artifacts

\Statex \textbf{Online posterior update}
\State Aggregate recent impressions $n_{i,s}$ and clicks $k_{i,s}$
\State Update posteriors:
\[
\alpha_{i,s}=\alpha_{0,i,s}+k_{i,s}, \quad
\beta_{i,s}=\beta_{0,i,s}+n_{i,s}-k_{i,s}
\]

\Statex \textbf{Serving}
\For{each request from segment $s$}
    \State Draw $\tilde{\theta}_{i,s}\sim\mathrm{Beta}(\alpha_{i,s},\beta_{i,s})$ for each candidate comment $i$
    \State Display the Top-$K$ comments by sampled value
\EndFor
\end{algorithmic}
\end{algorithm}

\FloatBarrier
\section{Additional Deployment Diagnostics}
\subsection{Exposure Concentration}
\label{app:exposure_concentration}

In Section~\ref{sec:discussion}, we discuss exposure concentration as one practical deployment consideration of LLM-initialized bandit serving. To quantify this effect, we compute title-level exposure concentration metrics for each prior configuration. For each title, we measure the number of uniquely exposed comments, the Herfindahl--Hirschman Index (HHI) of impression share across comments, and the Top-1 exposure share.

\begin{figure}[H]
    \centering
    \includegraphics[width=1\linewidth]{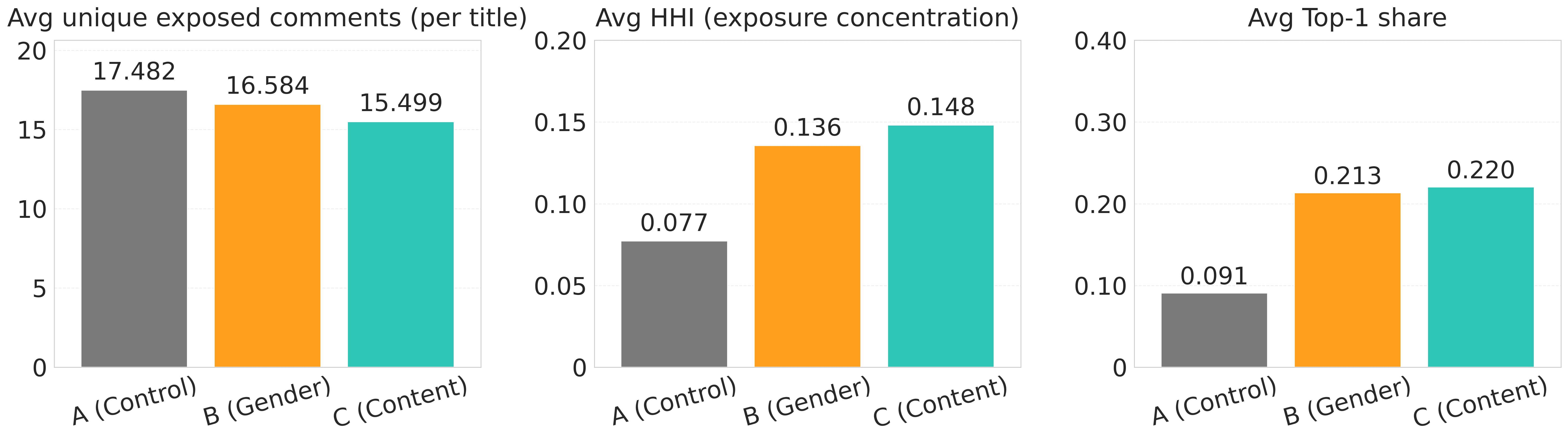}
    \caption{Exposure concentration summary across prior configurations. Bars report title-level averages of the number of uniquely exposed comments, the HHI of impression share, and the Top-1 exposure share. Relative to Variant~A, both Variant~B and Variant~C concentrate impressions on a smaller set of comments, indicating a more exploitative serving pattern with reduced exposure diversity.}
    \label{fig:exposure_concentration_summary}
\end{figure}

Figure~\ref{fig:exposure_concentration_summary} summarizes these metrics. Compared with the uniform-prior baseline, both LLM-prior variants expose fewer unique comments and concentrate impressions on a smaller set of comments. This concentration manifests the exploitation--diversity trade-off raised
in Section~5~\cite{kaminskas2016diversity,lathia2010temporal}.

\subsection{Popularity-Based Heterogeneity}
\label{app:popularity_heterogeneity}

As discussed in Section~\ref{sec:discussion}, title-level attractiveness can confound observed comment CTR and attenuate the measurable contribution of comment-level priors. We therefore conduct a supplementary popularity split to examine whether the effect of LLM-derived priors differs between popular and unpopular titles.

Table~\ref{tab:popularity_heterogeneity} reports CTR and CVR lift relative to Variant~A by title popularity group. Variant~B shows larger gains for unpopular titles than for popular titles in both CTR and CVR. Variant~C shows a different pattern: its CTR remains below the control in both popularity groups, while its CVR remains positive.

\begin{table}[H]
\centering
\caption{Lift relative to Variant~A by title popularity group. Variant~B shows larger gains for unpopular titles, whereas Variant~C shows a weaker and less uniform interaction with popularity.}
\label{tab:popularity_heterogeneity}
\begin{tabular}{lcc}
\toprule
\textbf{Group / Metric} & \textbf{Variant~B} & \textbf{Variant~C} \\
\midrule
CTR (Popular)     & +0.80\% & -6.15\% \\
CTR (Unpopular)   & +2.70\% & -4.44\% \\
CVR (Popular)     & +0.76\% & +1.42\% \\
CVR (Unpopular)   & +2.05\% & +1.07\% \\
\bottomrule
\end{tabular}
\end{table}
\balance
\end{document}